\documentclass[12pt]{iopart}
\usepackage{citesort}

\usepackage{inputenc}
\usepackage{graphicx}
\usepackage{subfigure}
\usepackage{color}
\usepackage{cite}
\usepackage{setspace}
\usepackage{cases}

\begin{document}

\title[Guided transport of ultracold gases of rubidium up to a room-temperature surface]{Guided transport of ultracold gases of rubidium up to a room-temperature dielectric surface}

\author{A L Marchant, S H\"{a}ndel, T P Wiles, S A Hopkins and S L Cornish}

\address{Department of Physics, University of Durham, South Road, Durham DH1 3LE, UK}
\ead{a.l.marchant@durham.ac.uk}
\begin{abstract}

We report on the guided transport of an atomic sample along an optical waveguide up to a room-temperature dielectric surface. The technique exploits a simple hybrid trap consisting of a single beam dipole trap positioned $\sim 125\,\mu$m below the field zero of a magnetic quadrupole potential. Transportation is realised by applying a moderate bias field ($<12\,$G) to displace the magnetic field zero of the quadrupole potential along the axis of the dipole trap. We use the technique to demonstrate that atomic gases may be controllably transported over $8\,$mm with negligible heating or loss. The transport path is completely defined by the optical waveguide and we demonstrate that, by aligning the waveguide through a super polished prism, ultracold atoms may be controllably delivered up to a predetermined region of a surface.

\end{abstract}

\maketitle


\section{Introduction}

Understanding the fundamental forces which govern the world around us has long been a challenge for the scientific community. Of particular interest is the search for a comprehensive explanation of gravity. Newton published his theory of gravitation in 1687~\cite{Principia} and since then attempts to experimentally verify its proposal have covered length scales from the astronomical~\cite{Kramer2004993} to the sub millimetre~\cite{PhysRevLett.86.1418}. The results of such experiments have not only fundamental significance, but also important technological implications.

Although these experiments seek to, and have imposed, increasingly strict bounds on fundamental forces there are still many open questions regarding the possibility of short-range corrections to gravity which extend beyond the Standard Model. Despite the electromagnetic, strong and weak forces all being well described by quantum field theories, the current description of gravity set out by Einstein's theory of general relativity, which reduces to Newtonian gravity on everyday length scales, breaks down in the quantum limit and as such, is currently excluded from the Standard Model. As a starting point, many experiments look for deviations from the expected inverse square law, with new forces instead being characterised by a Yukawa type potential of the form
\begin{equation}
U(r)=\frac{-G_{\rm N}m_1m_2}{r}(1+\alpha \textrm{e}^{-r/\lambda}),
\end{equation}
where $\alpha$ is the strength of the force and $\lambda$ its range. At the $1\,\mu$m level current experimental constraints permit these forces to be as large as $10^{10}$ times Newtonian gravity~\cite{PhysRevLett.102.171101}.

Attempts to measure the gravitational attraction between two masses have improved dramatically from the first experiments by Cavendish in 1798~\cite{Cavendish} and now vary widely in approach from superconducting gravity gradiometers~\cite{PhysRevLett.70.1195} and microcantilevers~\cite{PhysRevLett.90.151101} to planar oscillators~\cite{Nature.421.922} and torsion balance experiments~\cite{PhysRevLett.98.021101}. However in scaling down experiments to probe ever decreasing length scales a new, fundamental problem arises. Quantum electrodynamics predicts a macroscopic force between conductors, known as the Casimir force~\cite{Casimir.Proc.K.Ned}. This force vastly overwhelms the much weaker gravitational attraction between the test masses, such that experiments are forced to search for deviations between the theoretical and experimental Casimir forces. However, precisely calculating such Casimir forces for a specific macroscopic test mass near a surface is generally difficult~\cite{1367-2630-8-10-243}. In contrast the interaction between a single neutral atom and a plane surface is well understood~\cite{PhysRev.73.360, McLachlan:1963-1964:0026-8976:381} being characterised by the attractive Casimir-Polder potential,
\begin{numcases}{U_{\rm CP}=}
U_{\rm vdW}=-\frac{C_3}{z^3} & for $z < \lambda_{\rm opt}/2\pi $\\
U_{\rm ret}=-\frac{C_4}{z^4} & for $\lambda_{\rm opt}/2\pi < z < \lambda_{\rm T}$
\end{numcases}
where for longer length scales the $1/z^3$ form of the van der Waals potential, characterised by $C_3$, becomes $1/z^4$ due to retardation effects. This new regime is characterised by $C_4$ with the transition point between the two regimes determined by the wavelength  corresponding to the dominant excitation energy of the interacting atoms, $\lambda_{\rm opt}$ ~\cite{CPNature}. Further from the surface (larger than the thermal wavelength of photons, $\lambda_{\rm T}$) the interaction becomes dominated by the thermal fluctuation of the electromagnetic field~\cite{PhysRevA.70.053619}. For a rubidium atom and a dielectric surface with a refractive index of $\sim$1.5 these length scales are $\lambda_{\rm opt}/2\pi\approx0.12\,\mu$m and $\lambda_{\rm T}\approx 7.6\,\mu$m.

The inherent advantage of directly probing the atom-surface interaction has prompted the recent proposal of a new generation of experiments which aim to exploit the precision and control offered by atomic physics and ultracold quantum gases to push the measurement of short-range forces into a new regime~\cite{PhysRevD.68.124021, PhysRevA.75.063608, PhysRevLett.68.3432, NJP.8.237}. Indeed a number of proof-of-priniciple experiments have already utilised ultracold atomic gases to explore the short range van der Waals and Casimir-Polder potentials~\cite{PhysRevLett.77.1464, PhysRevLett.70.560, PhysRevLett.104.083201, PhysRevA.72.033610}. Nevertheless such experiments are in their infancy and considerable refinement is required before they become competitive with the classical `Cavendish style' experiments as a test of short-range gravitational forces.

Common to all these new atomic physics experiments is the need to controllably manipulate ultracold atoms near a room-temperature surface. Here we report the development of a new apparatus designed to study such atomic samples in close proximity to a super-polished surface of a dielectric Dove prism. In particular we describe and analyse a simple technique for the guided transport of an ultracold atomic gas along an optical waveguide. Our approach uses a hybrid optical and magnetic trap formed from a single optical dipole trap positioned $\sim 125\,\mu$m below the field zero of a magnetic quadrupole potential (see figure 1(a)-(c)). This simple hybrid trap allows us to position the sample precisely along the optical waveguide by the application of a moderate bias field to displace the field zero of the quadrupole potential. The transport path is completely defined by the optical waveguide and we demonstrate that, by aligning the waveguide through a super polished prism, ultracold atoms may be controllably delivered up to a predetermined region of a surface. We use this technique to demonstrate the loss of atoms due to the reduction of the trap depth as the trapping potential is brought up to the surface. In the future, such a technique will be employed to load atoms into a combined evanescent wave and optical lattice trap suitable for studying atom-surface interactions.

\begin{figure}
	\centering
		\includegraphics[width=0.9\textwidth]{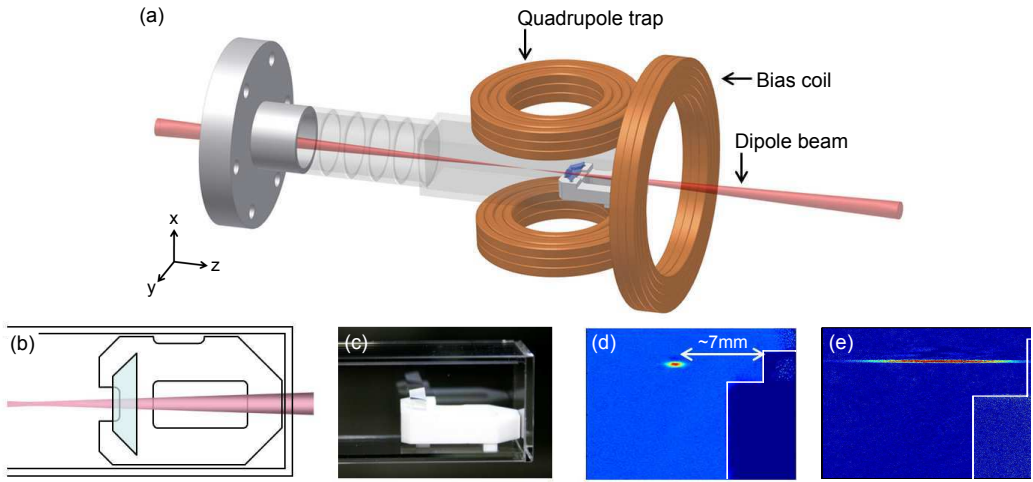}
	\caption{Experimental setup: (a) Trapping geometry near the surface. A single dipole beam is delivered through the back surface of the glass prism, focussing $3.5\,$mm from its front surface. Axial confinement along the beam is provided by a magnetic quadrupole field. A single coil positioned behind the prism is used to produce a bias field to shift the location of the quadrupole field zero along the beam direction. (b) Schematic of the glass cell, the Dove prism, the macor prism mount and the dipole beam from above. (c) Photograph of the prism within the glass cell. (d) False colour absorption image of atoms trapped $7\,$mm from the prism surface. (e) False colour absorption image of atoms in the waveguide without magnetic confinement.   }
	\label{fig:Figure1Setupandprism}
\end{figure}


\section{Modelling the trap potential}

\begin{figure}
	\centering
		\includegraphics[width=\textwidth]{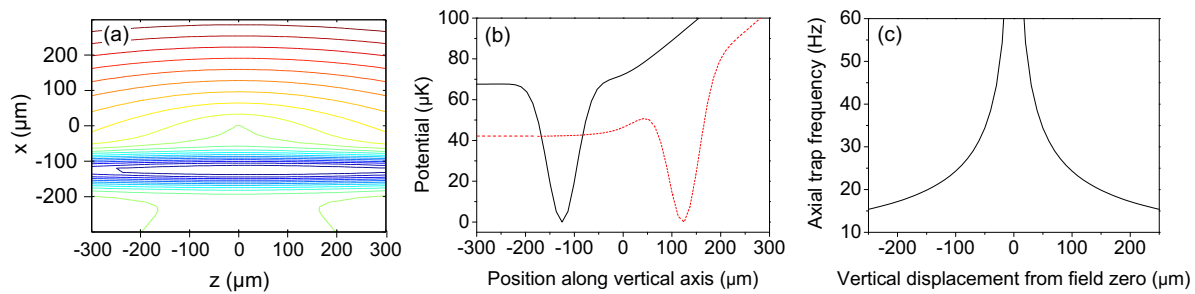}
	\caption{The hybrid trap: (a) Equipotential of the hybrid trap in the plane $y=0$ (see co-ordinate system in figure 1~(a)). (b) The potential obtained in the $x$ direction for a magnetically levitated trap with the dipole beam positioned $125\,\mu$m below (solid, black line) and $125\,\mu$m above (red, dashed line) the magnetic field zero. (c) Axial trap frequency as a function of vertical beam displacement from the magnetic field zero. }
	\label{fig:ModellingPotential}
\end{figure}

It is important to fully understand the hybrid potential formed by the combined magnetic and optical trap if atoms are to be translated up to and away from the surface in a controlled way. The total potential seen by the atoms has four contributions,
\begin{equation}
	U_{\rm{total}}=U_{\rm{dipole}}+U_{\rm{mag}}+U_{\rm{g}}+U_{\rm{CP}},
	\label{eg:totalpotential}
\end{equation}
where $U_{\rm dipole}$ is the optical dipole potential, $U_{\rm mag}$ is the magnetic potential (consisting of both the quadrupole and bias fields), $U_{\rm g}$ is the earth's gravitational potential and $U_{\rm CP}$ is the potential produced by the atom-surface interaction. This net potential is depicted in figure~\ref{fig:ModellingPotential}~(a) for a trap positioned far from the surface.

The optical contribution to the trap potential is modelled as a sum over any significant transitions from the ground state according to,
\begin{equation}
U_{\rm{dipole}}=3c^2\left(\sum_i\frac{\Gamma_i}{\Delta_i \omega_{0_i}^3}\right)\frac{P}{w^2(z)}\exp\left(-\frac{2r^2}{w^2(z)}\right).
\end{equation}
Here $\Delta_i$ is the laser detuning from the transition of frequency $\omega_{0_i}$ and natural linewidth $\Gamma_i$. $P$ is the power of the dipole beam propagating in the $z$ direction, $w(z)$ is the beam size, given by $w(z)=w_{0\rm M}[1+(z\lambda {\rm M}^2/\pi w_{0 \rm M}^2)^2]^{1/2}$ (where $w_{0\rm  M}$ is the 1/$e^2$ radius at the beam waist), and $r$ is the radial distance from the beam centre. In our case the dipole beam is derived from a $\lambda=1030\,$nm Yb disk laser and is focussed to a waist of $57\,\mu$m with an M$^2=1.06$ and Rayleigh range $z_R=9.9\,$mm. This beam is positioned 125$\,\mu$m below the field zero created by a magnetic quadrupole trap. The quadrupole gradient is usually set to approximately cancel gravity (30.6 G\,cm$^{-1}$ for $^{87}$Rb in the $F=1, m_F=-1$ state) hence the full trap depth is determined by the dipole beam alone. However, in an alternative beam position above the field zero, the trapping potential and the gravitational acceleration terms add to produce a tilted trap which lowers the potential barrier and hence trap depth as shown in figure~\ref{fig:ModellingPotential}~(b).

\subsection{Case I: Idealised transport in the hybrid trap}

In this hybrid trap configuration tight radial confinement is created by the dipole beam. Axial confinement along the beam is provided by the quadrupole field. (Without any magnetic confinement the cloud extends along the waveguide, centred around the beam waist, due to the low ($\sim2\,$Hz) axial trap frequency produced by the dipole beam alone as shown in figure~\ref{fig:Figure1Setupandprism}~(e).) The axial potential produced by the quadrupole field is harmonic for $z\ll x_{\rm offset}$ (where $x_{\rm offset}$ is the vertical separation between the beam and field zero) and linear otherwise. The harmonic trap frequency is determined by the quadrupole gradient and the vertical displacement of the beam from the field zero~\cite{PhysRevA.79.063631} as shown in figure~\ref{fig:ModellingPotential}~(c). Here the quadrupole gradient is sufficient to support atoms against gravity. In this hybrid trap the position of the atoms along the dipole beam is determined not by the beam waist but instead by the location of the quadrupole trap centre. The application of a horizontal bias field parallel to the dipole beam produces a shift of the field zero along the direction of the beam. In this way atoms can be transported along the length of the dipole beam and delivered up to a region of the prism surface determined solely by the optical alignment.

\subsection{Case II: Transport with an offset bias field}

In the idealised case the application of the bias field moves the field zero purely along the $z$-axis thus keeping the distance between the dipole beam and the field zero, and hence the axial trap frequency, constant. However, due to physical constraints of the apparatus it is unfeasible to position a pair of bias coils symmetrically about the beam. Instead we must use a single coil, displaced $17\,$cm in the $z$ direction from the field zero. Additionally the symmetry axis of the coil is offset $1.5\,$cm vertically. Theoretically accounting for the offset of the bias coil produces small deviations from the trapping expected for displacement along a gaussian beam. This is the result of the trajectory taken by the field zero as the bias field is increased due to a non-axial magnetic field component produced by the offset of the bias coil. As shown in figure~\ref{fig:ModellingPotential}~(c) any vertical displacement translates into a change in axial trap frequency. The theoretical model of this axial frequency change is shown, along with experimentally determined values, in section~\ref{LDH}. However, we stress that even in this non-ideal case the atoms are still transported along the dipole beam regardless of the exact path of the magnetic field zero.
%


\section{Production of ultracold gases near a dielectric surface}
\subsection{Loading the hybrid trap}

To prepare the atomic sample $^{87}$Rb atoms are loaded from a magneto-optical trap (MOT) into a quadrupole trap mounted on a motorised translation stage. Once loaded, this quadrupole trap is moved horizontally, transporting the atoms towards a second, static quadrupole trap (shown in figure~\ref{fig:Figure1Setupandprism}~(a)) into which the atoms are transferred. The atoms are then further cooled by forced RF evaporation resulting in a sample of $2.7\times~10^7$ atoms at a temperature of $32\,\mu$K. Further details of the apparatus are presented in~\cite{PhysRevA.83.053633, transport.paper}.

\begin{figure}
	\centering
		\includegraphics[width=0.9\textwidth]{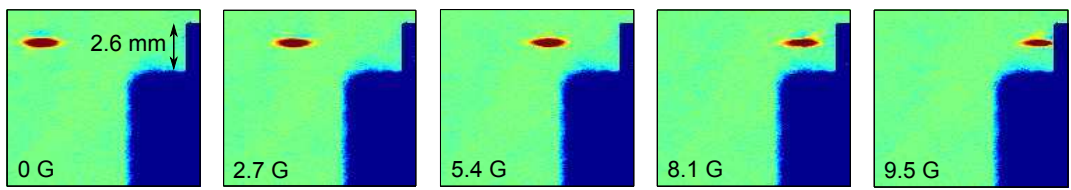}
	\caption{False colour absorption images of atoms approaching the surface. Bias fields applied range from $0\,$G to $9.5\,$G. To the right of each image the glass prism can be seen sharply defined on top of its macor mount (unfocussed edges).}
	\label{fig:MovingAtoms}
\end{figure}

To load the hybrid dipole trap the quadrupole gradient is relaxed from 192\,G\,cm$^{-1}$ down to 29.3\,G\,cm$^{-1}$ in $1\,$s. The combined potential produced by the dipole beam and quadrupole field results in trap frequencies of $\omega_r=2\pi\times480\,$Hz and $\omega_z=2\pi\times24\,$Hz at the beam waist. After loading, the cloud rapidly equilibrates to around $U_{\rm 0}/10$, where $U_{\rm 0}$ is the depth of the trap (this fraction we experimentally verify in section \ref{LDH}), roughly $7\,\mu$K. Further evaporation can then be performed by reducing the beam intensity. Throughout all experiments the dipole beam propagates through the rear anti-reflection coated face of the Dove prism, along the axis of the glass cell (see figure~\ref{fig:Figure1Setupandprism}). The waist is positioned $3.5\,$mm from the front super-polished surface. Initially the quadrupole trap centre is located $6.8\,$mm  from the prism surface, as shown in figure~\ref{fig:Figure1Setupandprism}(d). This geometry was chosen such that in moving up to the surface the cloud is always confined less than half a Rayleigh range from the beam waist, leading to minimal variation in the radial trapping potential. The cloud can be moved closer to and further from the prism by application of the bias field (where a positive displacement moves the cloud from the initial quadrupole trap centre, closer to the surface) as shown in figure \ref{fig:MovingAtoms}. The magnitude of this shift is given by,
\begin{equation}
z=\frac{B_0}{B'_z/2}
\end{equation}
where $B_0$ is the applied bias field and $B'_z$ is the quadrupole field gradient along the axis of the coils. For the (approximately) levitated potential the use of $-12\,$G$~\leq B_0~\leq 12\,$G produces displacements of $-8.2\,$mm $\leq z \leq 8.2\,$mm, on the order of a Rayleigh range ($z_R=9.9\,$mm).


\subsection{Speed of transport}

The maximum speed of the cloud transport without setting up sloshing in the trap is governed by the magnetic field gradient used to close off the trap along the axis of the dipole beam. Ideally the bias field ramp should be adiabatic to reduce any heating effects. This was experimentally verified by measuring the amplitude of the oscillation induced by the transportation. In order to map out the oscillation absorption images of the cloud were taken after quarter trap period time steps for atoms at temperatures of $1.5\,\mu$K and $7\,\mu$K. Figure~\ref{fig:4}~(a) shows that to complete a shift of $4.2\,$mm without setting up an oscillation requires a ramp time of $\sim\,2.5\,$s. As expected, this is independent of cloud temperature. In principle this time can be reduced with the use of a tighter field gradient at the expense of requiring a higher bias field.
%


\subsection{Characterisation of the hybrid trapping potential: A `single-shot' diagnostic}
\label{LDH}

With the flexibility to displace the trap centre anywhere along the beam, it is possible to use the trapped atoms to characterise the profile of the dipole beam. We use a `single-shot' diagnostic routine to measure important trap properties quickly and reliably. This approach is suitable when high radial trap frequencies are used, for example in single beam dipole traps. After a time of flight, $\tau_{\rm TOF}$, the width of the cloud, $\sigma_i$, is given by
\begin{equation}
\sigma_i^2=\left(\frac{k_{\rm{B}}T}{m\omega_i^2}\right)(1+\omega_i^2\tau^2_{\rm TOF}).
\end{equation}
In the limit that $\omega_i\tau_{\rm TOF} \gg 1$, i.e. in the case of the radial trap frequency, the cloud width after time of flight is governed only by the cloud temperature and not the frequency of the trap at release. Hence it is possible to determine the cloud temperature without prior knowledge of the trapping potential in a single shot. The axial trap frequency can then simply be calculated from the axial cloud size and the temperature as determined from the radial size. If the cloud is held in the trap sufficiently long before release such that the gas reaches full thermal equilibrium with the potential the radial trap frequency can also be derived from this measurement; knowledge of the dipole trap beam power together with the assumption that the cloud equilibrates to some fraction of the trap depth, $1/\eta$, (which we establish later in this section) allows the $1/e^2$ beam radius and hence the radial trap frequency to be determined from the temperature measurement. Such measurements are found to be in good agreement with the values obtained from parametric heating. This method allows the position of the beam waist to be located precisely and the distance between the magnetic field zero and the dipole beam to be determined over the full range of transport distances.

\begin{figure}
	\centering
		\includegraphics[width=0.9\textwidth]{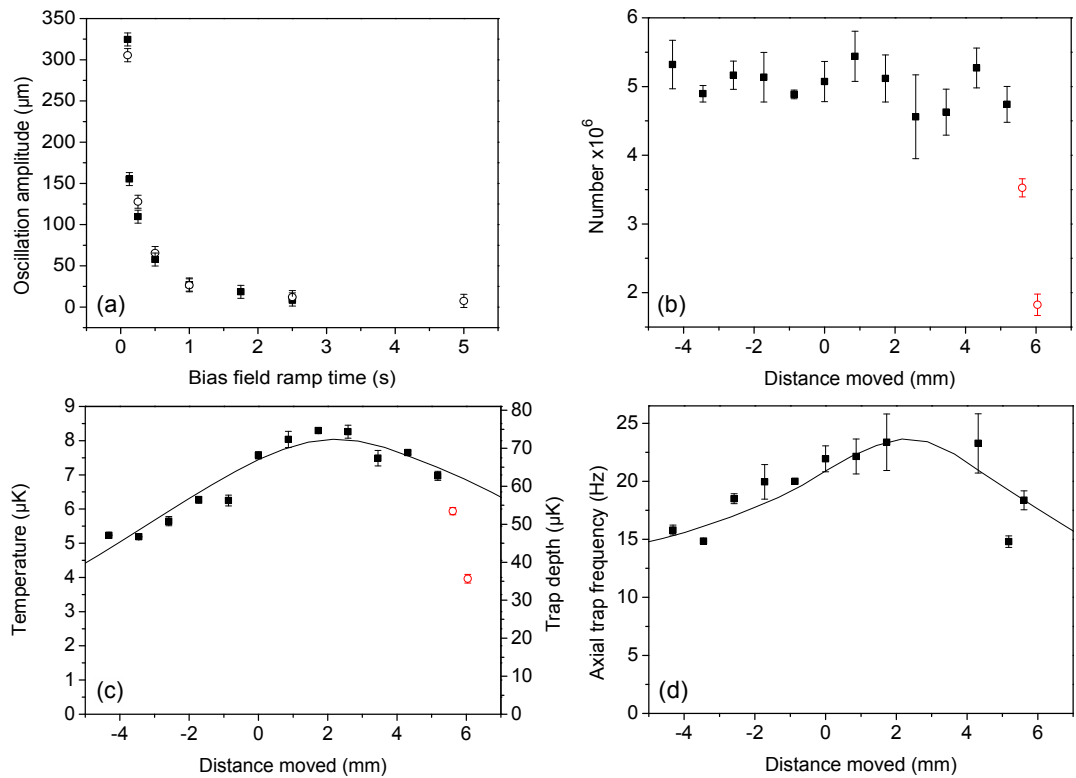}
		\caption{Characterisation of the transportation: (a) Oscillation set up by shifting the cloud $4.2\,$mm by displacement of the magnetic field zero in different lengths of time. Filled squares are for a cloud at $1.5\,\mu$K. Open circles are for a cloud at $7\,\mu$K. (b) Atom number as a function of horizontal trap shift caused by movement of the quadrupole field zero along the dipole beam. Red circles are for clouds sufficiently close to the surface that atoms are lost due to the atom-surface interaction opening up the trapping potential. (c) Vertical cloud temperature as a function of horizontal trap shift. Red circles are for clouds sufficiently close to the surface that atom loss from the cloud becomes a factor in the temperature. Solid line: Theoretical trap depth calculated from known dipole and quadrupole trap properties, accounting for an off axis bias field. Note the scale is $\times 9$ that of the experimental data indicating $\eta=9$ (see section~\ref{LDH}). (d) Experimentally determined axial trap frequency derived from the radial temperature and axial size of the cloud after time of flight. Solid line: Theoretical axial trap frequency calculated from known dipole and quadrupole trap properties, accounting for an off axis bias field. Displacements are with reference to the initial quadrupole trap centre location with no applied bias field.}
	\label{fig:4}
\end{figure}

Figure~\ref{fig:4}~(b) shows the effect of transport along the beam on atom number for a cloud initially allowed to come into thermal equilibrium with the trapping potential through evaporation. For shifts sufficiently far from the prism such that the atoms do not interact with the surface, there is no detectable atom loss. Closer to the prism there is a sharp drop in the atom number, the red (open) circles, as the atoms hit the surface (see section~\ref{section4}). The effect of the same transport on the cloud temperature is shown in figure~\ref{fig:4}~(c). The observed temperature change is due simply to the adiabatic compression and relaxation of the cloud as the radial trapping potential varies along the optical waveguide and not as a result of a heating mechanism associated with the motion of the cloud. Knowing the beam power at the trap and the beam waist (determined from parametric heating measurements) it is possible to model the trap depth along the waveguide. The solid line in figure~\ref{fig:4}~(c) demonstrates that the temperature data are consistent with a cloud in thermal equilibrium with the trapping potential for $\eta \approx 9$. Note again the two red (open) circles correspond to a shift sufficient for the surface potential to open up the trap, hence reducing the trap depth and leading to atom loss. As such, we do not expect agreement with the model potential for these points. To test explicitly for heating due to the transport, measurements were performed shifting a cloud cooled to 1/20 of the trap depth to suppress evaporation. Under such conditions some moderate heating was observed. For example, for a $1\,\mu$K cloud transported a round trip distance of $8\,$mm, heating rates of $\sim0.1\,\mu$Ks$^{-1}$ were observed for speeds that do not excite axial oscillations (see figure~\ref{fig:4}~(a)). However, the fact that the cloud remains in thermal equilibrium with the trapping potential (figure~\ref{fig:4}~(c)) and the absence of any atom loss (figure~\ref{fig:4}~(b)) demonstrates that this does not prohibit efficient transport of the atoms.

Analysis of the measured axial trap frequency moving along the beam confirms the theoretical prediction (solid line figure~\ref{fig:4}~(d)) that the vertical distance between the beam and the magnetic field zero does not remain constant when applying a bias field as a result of the bias coil's spatial offset from the beam axis. This leads to the small but measurable variation in trap frequency evident in figure~\ref{fig:4}~(d). However, this would be eliminated with a coil arrangement producing a field in the axial direction only.

\section{Loss due to the surface}
\label{section4}

\begin{figure}
	\centering
		\includegraphics[width=0.9\textwidth]{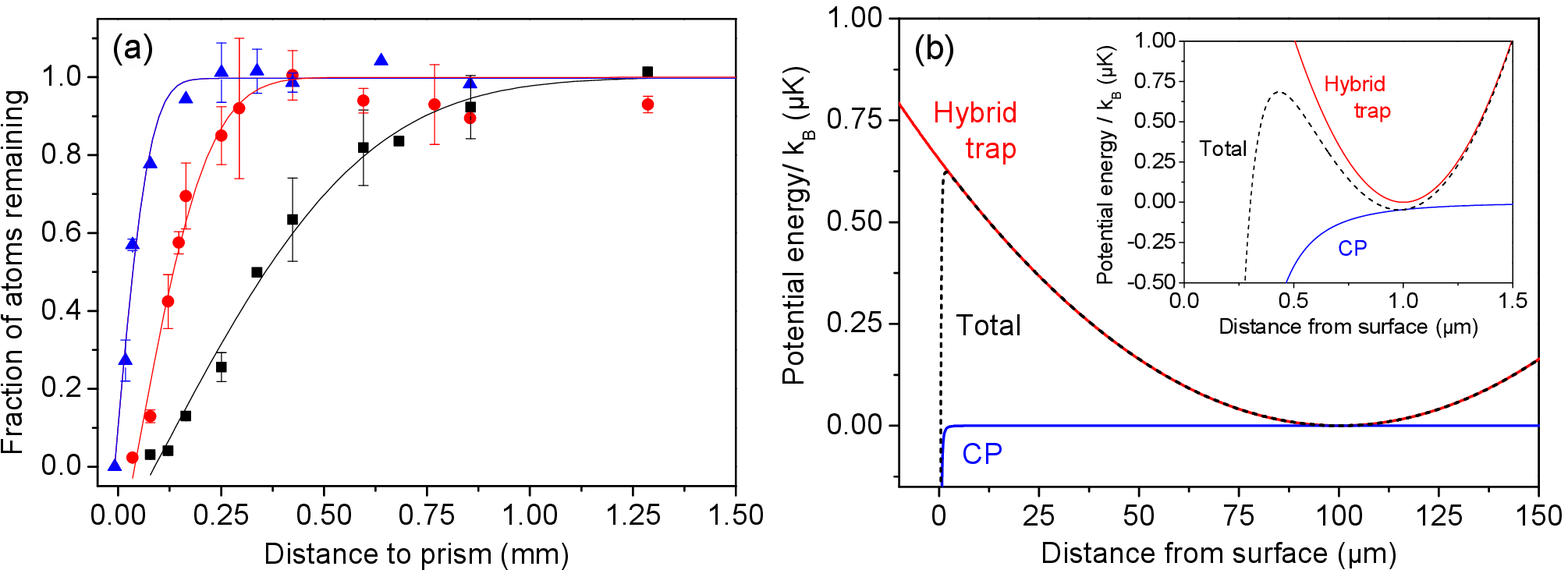}
	\caption{Atoms near the surface: (a) Atom loss as a function of distance from the prism surface for cloud temperatures of $7.0\,\mu$K (black squares), $1.4\,\mu$K (red circles) and $0.3\,\mu$K (blue triangles). Fitted lines are of the form $N=N_0\rm{erf}((z-z_0)/(\sqrt{2}\sigma)$) for $z>z_0$. (b) Total trapping potential resulting from the hybrid trap and Casimir Polder (CP) atom-surface potential when the trap is positioned $100\,\mu$m from the prism. Inset: The same resultant potential for a much tighter trap ($4.5\,$kHz) located only $1\,\mu$m from the surface.}
	\label{fig:Lossandpotentials}
\end{figure}

Far away from the surface the atomic cloud is unperturbed by displacement along the beam. However, once the distance of the trap centre from the surface becomes comparable to the axial cloud size atom loss is observed. This is a direct consequence of the strong attractive Casimir Polder potential leading to a finite trap depth along the optical waveguide (see figure~\ref{fig:Lossandpotentials}~(b)). Atoms whose energies are sufficient to escape over the finite barrier collide with the surface and are either adsorbed or reemitted at room temperature speeds. In this way the surface may be used as a knife for evaporative cooling~\cite{springerlink:10.1023/A:1026084606385}. Here we demonstrate how the atom loss caused by a controlled contact with the surface can be used to infer properties of the cloud such as size and temperature. We then proceed to produce a Bose-Einstein condensate by exploiting the evaporation of hot atoms at the surface.

The procedure used to probe the atom loss as the cloud approaches the surface is as follows. Atoms are loaded into the hybrid trap with no initial bias field and allowed to equilibrate to $7\,\mu$K . The bias field is then ramped to the necessary level in $5\,$s. Following this, the cloud is held at the shifted location for $50\,$ms (on the order of one axial trap period) before being shifted away from the surface a short distance and imaged. In the case of the colder cloud ($1.4\,\mu$K) an evaporation ramp in the dipole trap is first applied before the cloud is displaced along the beam. To produce the coldest cloud near the surface ($0.3\,\mu$K) an initial dipole evaporation stage is carried out and the cloud is shifted $0.8\,$mm from the prism before a further evaporation stage is employed to reach the final temperature.

The results of the measurement of loss due to the surface are shown in figure~\ref{fig:Lossandpotentials}~(a) as a function of the distance from the trap centre to the surface. Assuming harmonic axial confinement, it is possible to fit the observed atom loss as function of distance using an error function of the form
\begin{equation}
N=N_0\rm{erf}\left(\frac{z-z_0}{\sqrt{2}\sigma}\right),
\label{eq:width}
\end{equation}
for $z > z_0$ where $N$ is the number of atoms remaining at a given distance $z$ from the surface, $N_0$ is the initial atom number far from the surface and $\sigma$ is the cloud width. The offset $z_0$ accounts for the observation that, for the hotter clouds, complete atom loss occurs well before the trap centre reaches the prism. From the fit of equation~\ref{eq:width} to the loss data we find the cloud widths to be $400(20)\,\mu$m (for the $7.0\,\mu$K cloud), $140(20)\,\mu$m ($1.4\,\mu$K cloud) and $65(6)\,\mu$m ($0.3\,\mu$K cloud). Converting these widths to temperatures using $\sigma=(k_{\rm{B}}T/m\omega^2_z)^{1/2}$ we find reasonable agreement with time of flight measurements for the colder clouds. For the hotter clouds the anharmonic nature of the axial confining potential leads to deviations from the simple theoretical lineshape given by equation~(\ref{eq:width}).

The atom loss shown in figure~\ref{fig:Lossandpotentials}~(a) can be understood by consideration of the competing potentials as the hybrid trap is brought close to the surface. Figure~\ref{fig:Lossandpotentials}~(b) shows the result when the hybrid trap ($\omega_z=2\pi\times18$\,Hz) is positioned $100\,\mu$m from the surface. At these length scales the strong atom-surface potential causes a truncation of the harmonic trap resulting in a reduced trap depth. This leads to loss of hot atoms from the trap. The offset parameter $z_0$ can be understood by assuming that, due to finite signal to noise, all the atoms appear to be lost when the trap depth is reduced to 5\,--\,10\,\% of the initial cloud temperature. For example, for the cloud temperatures used in the experiment this reduction to $5\,\%$ occurs for trap-surface distances of $72\,\mu$m (for the $7.0\,\mu$K cloud), $32\,\mu$m ($1.4\,\mu$K cloud) and $15\,\mu$m ($0.3\,\mu$K cloud). These distances are broadly in agreement with the observed offsets in figure~\ref{fig:Lossandpotentials}~(a).

\begin{figure}
	\centering
		\includegraphics[width=0.45\textwidth]{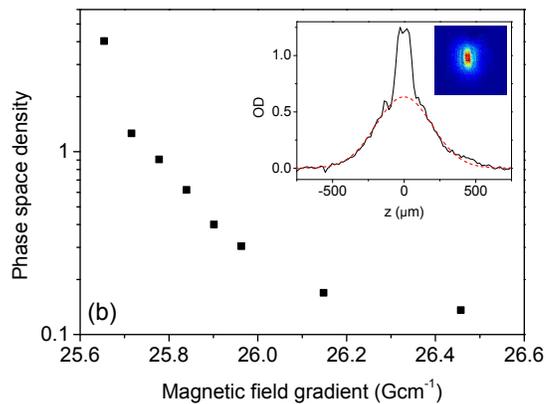}
	\caption{Cooling through the BEC transition using the surface. Phase space density as a function of magnetic field gradient for a fixed dipole beam intensity for a cloud initially positioned to be just in contact with the surface. Inset: Cross section through a cloud, initally at $1.5\,\mu$K and a phase space density of 2x$10^{-2}$, following evaporation to degeneracy using the surface. The false colour image shown is after $40\,$ms levitated time of flight.}
	\label{fig:figure6}
\end{figure}


This truncation of the trapping potential by the surface, if controlled, can also be used to cool the trapped atoms. Relaxation of the field gradient reduces the trapping along the dipole beam thus allowing atoms to extend outwards from the trap centre, towards the prism, coming into contact with the $300\,$K surface. The resulting atom loss due to the surface is very similar to the application of an RF knife to selectively remove the most energetic atoms. Carrying out evaporation in this way allows the BEC transition to be reached (see figure~\ref{fig:figure6}) without the need to reduce the dipole intensity further than a first evaporation stage. However, due to the limited and hence inefficient evaporation surface produced by the trapping geometry it was not possible to evaporate to degeneracy solely using this surface technique.


\section{Conclusion and outlook}

We have demonstrated the successful transport of an atomic sample along an optical waveguide up to a room-temperature dielectric surface. The technique exploits a simple hybrid trap consisting of a single beam dipole trap positioned $\sim125\,\mu$m below the field zero of a magnetic quadrupole potential. Transportation is realised by applying a moderate bias field ($<12\,$G) to displace the magnetic field zero of the quadrupole potential along the axis of the dipole trap. We use the technique to demonstrate that atomic gases may be controllably transported over $8\,$mm with negligible heating or loss. This distance can be readily extended using a slightly modified coil and beam geometry. The transport path is completely defined by the optical waveguide and we demonstrate that, by aligning the waveguide through a super polished  Dove prism, ultracold atoms may be controllably delivered up to a predetermined region of a surface. Upon approaching the surface we observe strong atom loss once the distance of the trap centre from the surface becomes comparable to the axial cloud size. This is simply a direct consequence of the presence of the surface leading to a reduced trap depth along the optical waveguide. Such loss is akin to evaporative cooling and, indeed, we demonstrate how the effect can be utilized to cool an atomic gas through the transition to a Bose-Einstein condensate.

The next phase of our experiment is to use the guided transport technique to load the atoms into an optical trap in the vicinity of the surface which is suitable for studying atom-surface interactions. Typically such a trap needs to be tightly confining~\cite{PhysRevLett.104.083201, PhysRevA.79.013409, PhysRevLett.92.173003, Nature.1.57, PhysRevA.80.021602} so that the spatial extent of the atomic cloud becomes comparable to the range of the Casimir Polder potential. This allows the atomic gas to be positioned much closer ($<10\,\mu$m) to the surface.  For example, the inset to figure 5 (b) shows the resulting trapping potential for a $4.5\,$kHz harmonic trap positioned at a distance of $1\,\mu$m from the surface. At such length scales the Casimir Polder potential leads not only to a reduction in the trap depth but also to a change in the trapping frequency, an observable which can itself be useful~\cite{PhysRevLett.98.063201}. The use of a Dove prism in our experiment permits the addition of evanescent wave potentials at the surface through the total internal reflection of light within the prism. In particular, the use of a blue-detuned laser will create a repulsive barrier close to the surface (an evanescent wave atomic mirror\cite{Kasevich:90, Christ1994211, Seifert1994566, PhysRevLett.77.1464, PhysRevLett.88.250404, 0953-4075-39-22-009, springerlink:10.1007/s00340-009-3564-2}) that will prevent the observed loss as the atoms are transported up to the surface. Previous work has used purely magnetic transport to bring atoms close to such a barrier~\cite{1464-4266-5-2-374}. In the scheme presented here, the use of an optical waveguide has the advantage that the transport path is insensitive to stray magnetic fields as the atoms are always guided along the beam towards the target point on the surface. The technique also has the advantage of simplicity over other transport schemes involving optical lattices~\cite{1367-2630-8-8-159, PhysRevA.79.013409}. Once the transportation up to the surface is complete, atoms will be loaded into a tight surface trap created either by the addition of a second red-detuned evanescent field~\cite{0953-4075-24-14-009, Desbiolles1996540, PhysRevA.61.023608, PhysRevLett.90.173001} or by the reflection of a red-detuned laser incident at a shallow angle on the surface from the vacuum side. The latter approach can be used to generate a 1D optical lattice close to the surface~\cite{PhysRevA.80.021602} which can potentially be tailored in order to test a number of novel schemes to measure the atom-surface potential, including interferomtery in a double-well potential ~\cite{PhysRevA.82.032104} and the study of Bloch oscillations~\cite{PhysRevLett.106.038501}. To ultimately probe short range corrections to gravity will require a detailed comparison of the measured and theoretical Casimir-Polder potentials. Pushing ultracold atom experiments to the precision to be competitive with traditional approaches will be challenging and will undoubtedly require the development of further techniques to transport and manipulate atomic gases close to surfaces.


\section*{Acknowledgments}
We acknowledge support from the UK Engineering and Physical Sciences Research Council (EPSRC grant EP/F002068/1) and the European Science Foundation within the EUROCORES Programme EuroQUASAR (EPSRC grant EP/G026602/1). SLC acknowledges the support of the Royal Society.



\section*{References}

\providecommand{\newblock}{}


\begin{thebibliography}{10}
\expandafter\ifx\csname url\endcsname\relax
  \def\url#1{{\tt #1}}\fi
\expandafter\ifx\csname urlprefix\endcsname\relax\def\urlprefix{URL }\fi
\providecommand{\eprint}[2][]{\url{#2}}

\bibitem{Principia}
Newton I 1687 {\em Philosophi\ae Naturalis Principia Mathematica\/} (London:
  The Royal Society)

\bibitem{Kramer2004993}
Kramer M, Backer D, Cordes J, Lazio T, Stappers B and Johnston S 2004 {\em New
  Astronomy Reviews\/} {\bf 48} 993 -- 1002

\bibitem{PhysRevLett.86.1418}
Hoyle C~D, Schmidt U, Heckel B~R, Adelberger E~G, Gundlach J~H, Kapner D~J and
  Swanson H~E 2001 {\em Phys. Rev. Lett.\/} {\bf 86} 1418--1421

\bibitem{PhysRevLett.102.171101}
Masuda M and Sasaki M 2009 {\em Phys. Rev. Lett.\/} {\bf 102} 171101

\bibitem{Cavendish}
Cavendish H 1798 {\em Roy. Soc. Phil. Trans.\/} {\bf 88} 469

\bibitem{PhysRevLett.70.1195}
Moody M~V and Paik H~J 1993 {\em Phys. Rev. Lett.\/} {\bf 70} 1195--1198

\bibitem{PhysRevLett.90.151101}
Chiaverini J, Smullin S~J, Geraci A~A, Weld D~M and Kapitulnik A 2003 {\em
  Phys. Rev. Lett.\/} {\bf 90} 151101

\bibitem{Nature.421.922}
Long J~C, Chan H~W, Churnside A~B, Gulbis E~A, Varney M~C~M and Price J~C 2003
  {\em Nature\/} {\bf 421} 922

\bibitem{PhysRevLett.98.021101}
Kapner D~J, Cook T~S, Adelberger E~G, Gundlach J~H, Heckel B~R, Hoyle C~D and
  Swanson H~E 2007 {\em Phys. Rev. Lett.\/} {\bf 98} 021101

\bibitem{Casimir.Proc.K.Ned}
Casimir H 1948 {\em Proc. K. Ned. Akad. Wet\/} {\bf 51} 793--795

\bibitem{1367-2630-8-10-243}
Lambrecht A, Neto P~A~M and Reynaud S 2006 {\em New J. Phys.\/} {\bf 8} 243

\bibitem{PhysRev.73.360}
Casimir H~B~G and Polder D 1948 {\em Phys. Rev.\/} {\bf 73} 360--372

\bibitem{McLachlan:1963-1964:0026-8976:381}
McLachlan A 1963-1964 {\em Mol. Phys.\/} {\bf 7} 381--388(8)

\bibitem{CPNature}
Casimir H~B~G and Polder D 1946 {\em Nature\/} {\bf 158} 787--788

\bibitem{PhysRevA.70.053619}
Antezza M, Pitaevskii L~P and Stringari S 2004 {\em Phys. Rev. A\/} {\bf 70}
  053619

\bibitem{PhysRevD.68.124021}
Dimopoulos S and Geraci A~A 2003 {\em Phys. Rev. D\/} {\bf 68} 124021

\bibitem{PhysRevA.75.063608}
Wolf P, Lemonde P, Lambrecht A, Bize S, Landragin A and Clairon A 2007 {\em
  Phys. Rev. A\/} {\bf 75} 063608

\bibitem{PhysRevLett.68.3432}
Sandoghdar V, Sukenik C~I, Hinds E~A and Haroche S 1992 {\em Phys. Rev.
  Lett.\/} {\bf 68} 3432--3435

\bibitem{NJP.8.237}
Onofrio R 2006 {\em New J. Phys.\/} {\bf 8} 237

\bibitem{PhysRevLett.77.1464}
Landragin A, Courtois J~Y, Labeyrie G, Vansteenkiste N, Westbrook C~I and
  Aspect A 1996 {\em Phys. Rev. Lett.\/} {\bf 77} 1464--1467

\bibitem{PhysRevLett.70.560}
Sukenik C~I, Boshier M~G, Cho D, Sandoghdar V and Hinds E~A 1993 {\em Phys.
  Rev. Lett.\/} {\bf 70} 560--563

\bibitem{PhysRevLett.104.083201}
Bender H, Courteille P~W, Marzok C, Zimmermann C and Slama S 2010 {\em Phys.
  Rev. Lett.\/} {\bf 104} 083201

\bibitem{PhysRevA.72.033610}
Harber D~M, Obrecht J~M, McGuirk J~M and Cornell E~A 2005 {\em Phys. Rev. A\/}
  {\bf 72} 033610

\bibitem{PhysRevA.79.063631}
Lin Y~J, Perry A~R, Compton R~L, Spielman I~B and Porto J~V 2009 {\em Phys.
  Rev. A\/} {\bf 79} 063631

\bibitem{PhysRevA.83.053633}
H\"andel S, Wiles T~P, Marchant A~L, Hopkins S~A, Adams C~S and Cornish S~L
  2011 {\em Phys. Rev. A\/} {\bf 83} 053633

\bibitem{transport.paper}
H\"{a}ndel S, Marchant A~L, Wiles T~P, Hopkins S~A and Cornish S~L 2011 {\em
  \rm{arXiv:1109.5340v1}\/}

\bibitem{springerlink:10.1023/A:1026084606385}
Harber D~M, McGuirk J~M, Obrecht J~M and Cornell E~A 2003 {\em J. Low Temp. Phys.\/}
  {\bf 133}(3) 229--238

\bibitem{PhysRevA.79.013409}
Sorrentino F, Alberti A, Ferrari G, Ivanov V~V, Poli N, Schioppo M and Tino G~M
  2009 {\em Phys. Rev. A\/} {\bf 79} 013409

\bibitem{PhysRevLett.92.173003}
Rychtarik D, Engeser B, N\"agerl H~C and Grimm R 2004 {\em Phys. Rev. Lett.\/}
  {\bf 92}(17) 173003

\bibitem{Nature.1.57}
Schumm T, Hofferberth S, Andersson L~M, Wildermuth S, Groth S, Bar-Joseph I,
  Schmiedmayer J and Kr\"uger P 2005 {\em Nature Phys.\/} {\bf 1} 57

\bibitem{PhysRevA.80.021602}
Gillen J~I, Bakr W~S, Peng A, Unterwaditzer P, F\"olling S and Greiner M 2009
  {\em Phys. Rev. A\/} {\bf 80}(2) 021602

\bibitem{PhysRevLett.98.063201}
Obrecht J~M, Wild R~J, Antezza M, Pitaevskii L~P, Stringari S and Cornell E~A
  2007 {\em Phys. Rev. Lett.\/} {\bf 98} 063201

\bibitem{Kasevich:90}
Kasevich M~A, Weiss D~S and Chu S 1990 {\em Opt. Lett.\/} {\bf 15} 607--609

\bibitem{Christ1994211}
Christ M, Scholz A, Schiffer M, Deutschmann R and Ertmer W 1994 {\em Opt.
  Commun.\/} {\bf 107} 211 -- 217

\bibitem{Seifert1994566}
Seifert W, Kaiser R, Aspect A and Mlynek J 1994 {\em Opt. Commun.\/} {\bf 111}
  566 -- 576

\bibitem{PhysRevLett.88.250404}
Savalli V, Stevens D, Est\`eve J, Featonby P~D, Josse V, Westbrook N, Westbrook
  C~I and Aspect A 2002 {\em Phys. Rev. Lett.\/} {\bf 88}(25) 250404

\bibitem{0953-4075-39-22-009}
Perrin H, Colombe Y, Mercier B, Lorent V and Henkel C 2006 {\em J. Phys. B: At.
  Mol. Phys.\/} {\bf 39} 4649

\bibitem{springerlink:10.1007/s00340-009-3564-2}
Bender H, Courteille P, Zimmermann C and Slama S 2009 {\em Appl. Phys. B:
  Lasers and Opt.\/} {\bf 96}(2) 275--279

\bibitem{1464-4266-5-2-374}
Colombe Y, Kadio D, Olshanii M, Mercier B, Lorent V and Perrin H 2003 {\em J.
  Opt. B: Quant. Semiclass. Opt.\/} {\bf 5} S155

\bibitem{1367-2630-8-8-159}
Schmid S, Thalhammer G, Winkler K, Lang F and Denschlag J~H 2006 {\em New J.
  Phys\/} {\bf 8} 159

\bibitem{0953-4075-24-14-009}
Ovchinnikov Y~B, Shul'ga S~V and Balykin V~I 1991 {\em J Phys. B: At. Mol.
  Phys.\/} {\bf 24} 3173

\bibitem{Desbiolles1996540}
Desbiolles P and Dalibard J 1996 {\em Opt. Commun.\/} {\bf 132} 540 -- 548

\bibitem{PhysRevA.61.023608}
Barnett A~H, Smith S~P, Olshanii M, Johnson K~S, Adams A~W and Prentiss M 2000
  {\em Phys. Rev. A\/} {\bf 61}(2) 023608

\bibitem{PhysRevLett.90.173001}
Hammes M, Rychtarik D, Engeser B, N\"agerl H~C and Grimm R 2003 {\em Phys. Rev.
  Lett.\/} {\bf 90}(17) 173001

\bibitem{PhysRevA.82.032104}
Chwede\ifmmode~\acute{n}\else \'{n}\fi{}czuk J, Pezz\'e L, Piazza F and Smerzi
  A 2010 {\em Phys. Rev. A\/} {\bf 82} 032104

\bibitem{PhysRevLett.106.038501}
Poli N, Wang F~Y, Tarallo M~G, Alberti A, Prevedelli M and Tino G~M 2011 {\em
  Phys. Rev. Lett.\/} {\bf 106} 038501

\end{thebibliography}

\end{document}